# The Puzzling Mutual Orbit of the Binary Trojan Asteroid (624) Hektor

F. Marchis[1,2,*], J. Durech[3], J. Castillo-Rogez[4], F. Vachier[2], M. Cuk[1], J. Berthier[2], M.H. Wong[5], P. Kalas[5], G. Duchene[5,6], M. A. van Dam[7], H. Hamanowa[8], M. Viikinkoski[9]

[1]Carl Sagan Center at the SETI Institute, Mountain View, California, 94043 USA.
[2]IMCCE-Obs de Paris, 75014 Paris, France.
[3]Astronomical Institute, Faculty of Mathematics and Physics, Charles University, Prague, Czech Republic
[4]Jet Propulsion Laboratory, California Institute of Technology, California, USA.
[5]Department of Astronomy at UC Berkeley, 94720 California, USA.
[6]Institut de Planétologie et d'Astrophysique de Grenoble, 38041 Grenoble, France.
[7]Flat Wavefronts, Christchurch 8140, New Zealand.
[8]Hamanowa Observatory, Motomiya, Fukushima 969-1204, Japan.
[9]Tampere University of Technology, 33101 Tampere, Finland

\* Corresponding author. E-mail: fmarchis@seti.org

**Abstract:** Asteroids with satellites are natural laboratories to constrain the formation and evolution of our solar system. The binary Trojan asteroid (624) Hektor is the only known Trojan asteroid to possess a small satellite. Based on W.M. Keck adaptive optics observations, we found a unique and stable orbital solution, which is uncommon in comparison to the orbits of other large multiple asteroid systems studied so far. From lightcurve observations recorded since 1957, we showed that because the large $R_{eq}$=125-km primary may be made of two joint lobes, the moon could be ejecta of the low-velocity encounter, which formed the system. The inferred density of Hektor's system is comparable to the L5 Trojan doublet (617) Patroclus but due to their difference in physical properties and in reflectance spectra, both captured Trojan asteroids could have a different composition and origin.

1. INTRODUCTION

As of today ~200 asteroids are known to possess one or several satellites across all populations of small solar system bodies (Richardson and Walsh, 2006), from the near-earth asteroids (Pravec et al. 2006), to the asteroid main-belt (Marchis et al. 2008) and beyond (Noll et al. 2008). Their existence and the study of their mutual orbits lead to significant constraints on the formation of our solar system through the determination of the mass, densities, mass ratio and long term evolution of the system. In the Jupiter Trojan population, (617) Patroclus is the only doublet asteroid systems, made of two comparable-sized components, which has been imaged and studied so far (Marchis et al. 2006b). We report in this work on the shape and interior of the L4 multiple Trojan (624) Hektor and on the analysis of the orbit of its moon based on our AO observations and additional datasets. We discuss the internal structure of the primary in comparison with the one of another binary Trojan asteroid (617) Patroclus.

2. DATA ANALYSIS
2.1 Mutual Orbit

Using the potential of the W.M. Keck Laser Guide Star (LGS) system (Bouchez et al. 2004) and its adaptive optics system (Wizinowich et al. 2000), we conducted a search survey (Marchis al. 2006c) and found one satellite (Marchis et al. 2006a) around the Jupiter-Trojan asteroid (624) Hektor on July 16 2006 by directly imaging the asteroid in the near-infrared wavelength range. The epoch of observations and the relative positions and brightness of the satellite are listed in Table 1. The first detection was made using the $K_p$ ($\lambda_c$=2.2 μm) filter, with the satellite then located at 0.362" from the primary with a difference in brightness of 4.2 mag. It was also marginally visible on frames recorded using in the H broadband filter ($\lambda_c$=1.6 μm). We made follow-up observations on 16 occasions since that date using LGS or the natural guide star system (NGS) at W.M. Keck Observatory and also using the Very Large Telescope, both equipped with adaptive optics (AO). Due to the faintness of the moon and its proximity to the primary, only 12 W.M. Keck observations recorded from July 2006 to November 2011 (Figure 1) revealed its presence. The primary is elongated and resolved in the elongated direction with an angular diameter of ~0.110". Assuming the same surface composition (hence a similar albedo in the near-infrared) we estimate the moon to have a diameter of 12 ± 3 km. This discovery is the first evidence on the existence of satellites around Jupiter-Trojan asteroids. The presence of one small moon around the large Trojan asteroid (911) Agamemnon (Timerson et al. 2013) has been suggested by a secondary occultation event.

To derive the mutual orbit of the system, we used Genoide-ANIS, a genetic-based algorithm specifically designed for the analysis of binary asteroid systems (Vachier et al. 2012), which fits the orbital parameters of the orbit, the orientation and size of the primary, and the degree two gravity harmonics ($J_2$=-$C_{20}$). This algorithm was successfully applied on well-constrained binary systems such as 22 Kalliope (Vachier et al. 2012) and 87 Sylvia (Marchis et al. 2013). Table 1 summarizes the parameters extracted from this algorithm. The solution is accurate with a fitness function *fp* of 10 milli-arcsec, i.e., of the same order as the astrometric accuracy on the satellite's positions (~15 milli-arsec). Precession effects due to the elongated shape of the primary ($J_2$ = 0.15) are clearly detected. The orbit of the secondary (Table 2) is different from those derived in the case of other large (*D* > 100 km) binary asteroid systems (Marchis et al. 2008) with unique features such as a significant eccentricity (*e*~0.3) and an inclination of ~50 deg with respect to the equator of the primary.

**2.2 Shape of the Primary**

To derive the shape models of Hektor together with the rotation period and the spin pole position, we used 75 lightcurves observed from published observations taken from 1957 to 1991 (Dunlap & Gehrels, 1969; Hartmann & Cruikshank, 1978; Detal et al. 1994; Hainaut-Rouelle et al. 1995), and recent lightcurves collected by our group from 2006 to 2011. We also included sparse-in-time photometry obtained by US Naval Observatory and Catalina Sky Survey. (624) Hektor's primary elongation has been known since the 1980's because of its large brightness variations (Dunlap and Gehrels, 1969), which suggest a bilobed nature (Hartmann and Cruikshank, 1978). From this dataset, we were able to reconstruct a convex shape model with two possible pole positions. The solutions are consistent with the results of Kaasalainen, 2002, who used only 19 lightcurves form 1957-1991. However since only one pole position is consistent with the AO data, we are now able to remove the ambiguity and select the single consistent solution. In addition to the convex model, we also made a bilobed and a close-binary model of Hektor. All three models shown in Figure 2A-C have very similar fitting quality for the photometric data.

In principle, AO data can be used together with lightcurves to reconstruct a more detailed non-convex model of an asteroid. However, in the case of Hektor, the bulk of the AO images were not of sufficient resolution to reveal any details of the shape. They served mostly for discriminating the pole solution and for scaling the model. However, some of the best images in Fig. 1 recorded with the Keck AO system suggest a dual structure for the primary, supporting the bilobed shape model.

Because (624) Hektor is farther from the Sun and the Earth than main belt asteroids, the solar phase angle cannot reach large values and shadowing effects are not important even for a highly non-convex body. Using the Hubble Space Telescope/Fine Guider Sensor (HST/FGS) S-curves available in HST archive as an additional source of information, we found out that the binary model gave the worse fit, so it was rejected, while both convex and bilobed models exhibited similar fits. Additionally, an occultation recorded in January 24, 2008, by Hektor of a 10 mag star visible from Japan suggested that Hektor could be a binary. Because the observation of the double chord was made visually through a 20cm telescope during twilight, we do not consider this result to be sufficiently reliable to make this claim.

The ecliptic coordinates of the spin axis are (332 deg, -27 deg) and (332 deg, -32 deg) for the convex and bilobed models, respectively. Because the 1-σ uncertainty on the pole position is about 5 deg, the poles are virtually the same. The rotation period for both shape models is P = 6.920509 ± 0.000002 h. The equivalent diameter, or the diameter of a sphere with the same volume than the asteroid, is in the same range (within the errors): 270 ± 22 km for the convex model and 256 ± 12 km for the bilobed model very close to the Genoide-ANIS solution (224 ± 34 km). We adopted an equivalent diameter of 250 ± 26 km for Hektor's primary.

We calculated the theoretical $J_2$ of those models assuming a homogenous distribution of material in the interior which is ~0.08 and ~0.10 for the convex and bilobed shape mode. Since the polar flattening is not well constrained, $J_2$ is not well determined. Genoide-ANIS gives a dynamical $J_2$ marginally larger ($J_2 \sim 0.15$). This measurement is compatible with Hektor's primary being made of two joint components which could be inhomogeneous, possibly differentiated.

**2.3 Long-Term Dynamics of Hektor's Satellite**

The orbits of satellites around asteroids are perturbed by a number of effects, including the non-spherical shapes of the two bodies, solar gravity, solar radiation pressure and its derivatives, gravity of any additional satellites, and (during close encounters) the gravity of other planets and asteroids (Ćuk and Nesvorný, 2010, and references therein). In general, perturbations from the components' shapes are dominant for close satellites (e.g. NEA and MBA binaries, regular satellites of the giant planets) while solar perturbations tend to become more important for distant satellites (e.g. Earth's Moon, irregular satellites of the giant plants, wide TNO binaries). Radiation forces typically affect only small binaries (with primaries smaller than a few km across), while close encounters affect only binaries on planet-crossing orbits (like some NEA and TNO binaries).

If a lone satellite's observed orbit is clear of the primary's Roche limit (typically at 2-3 primary radii) and stays well within the primary's Hill radius (typically at hundreds of primary radii for main belt asteroids), it is generally stable in the short-term (Nicholson et al. 2008). In the long term, any of the above-mentioned perturbations could evolve the initial orbit into one that may hit the primary or escape the system, so a more in-depth analysis is needed to determine a satellite's long-term orbital stability.

We modeled long-term evolution of the Hektor binary using a symplectic integrator which treats rotating Hektor as an oblate body and includes solar perturbations (for details see Ćuk and Gladman, 2009). We find that the orbit of Hektor's satellite is likely to be stable in the long term, based both on analytical and numerical considerations. The orbital period is an order of magnitude longer than the spin period of Hektor, with the period ratio of 10.28. Also, the nominal orbit avoids powerful resonances with the rotation of the primary at 10 and 21/2 times the primary spin period (Petit, 1997). These resonances put serious constraints on both the present orbit of the satellite, and any past orbital evolution. However, if the orbit is significantly less eccentric than the nominal solution ($e = 0.3$), the robustness of these constraints is weakened.

Large inclination and eccentricity may appear unstable, but this is not the case as orbital perturbations are dominated by Hektor's effective oblateness (resulting from fast rotation of a triaxial body). Unlike solar ("Kozai" Nicholson et al. 2008) perturbations, the quadrupole component of planetary oblateness does not allow the exchange of angular momentum between eccentricity and inclination, allowing for orbits at all inclinations to have relatively unchanging elements (Murray and Dermott, 1999). As the precession of the mutual orbit plane is primarily around Hektor's equator, the mutual orbit cycles between prograde and retrograde relative to the heliocentric motion, while maintaining constant inclination with respect to the equator (Figure 3F).

Since the precession periods of the mutual orbit are of the same order as the heliocentric mean motion, there are some small-amplitude cycles in the orbital elements that are associated with semi-secular arguments (Figure 3A-G). Considering that its orbit is not well known, it is possible that it may be in an actual semi-secular resonance (for example, if the nodal precession period is equal to one half of Jupiter's year). However, phase space taken by such resonances is small and, in all likelihood, no such resonances are present. The longer-term behavior of the orbit is stable and unchanging (Figure 3H-I).

On longer timescales, non-conservative forces need to be taken into account. These include tides and the BYORP (Yarkovsky-O'Keefe-Radzievski-Paddack) radiation effect (Ćuk & Burn, 2005). Not only would BYORP effect be weaker for Hektor than among small NEA binaries, but the large eccentricity of the mutual orbit all but guarantees that the secondary is in chaotic rotation (Ćuk and Nesvorný, 2010), which would suppress BYORP. Using classical tidal expressions (Murray and Dermott, 1999) with tidal quality factor Q=100, Love number $k_2$=0.0001 and R=129 km for Hektor and R=7.5 km for the satellite (with the same density), the timescale for the tidal evolution should be about 30 Gyr, or longer than the age of the Solar System. As the satellite's collisional lifetime is shorter, Hektor may be a relatively unperturbed system that has not evolved significantly through tides, with the large eccentricity and inclination of the satellite being primordial.

## 3. INTERPRETATION OF (624) HEKTOR's DENSITY

We found an average bulk density of Hektor's system ($\rho$ = 1.0 ± 0.3 g/cm$^3$) that is very close to the one of (617) Patroclus-Menoetius, a doublet binary Trojan asteroid system in Jupiter's L5 cloud (Marchis et al. 2006b) made of two components of 106 and 98 km in diameter (Mueller et al. 2010). Both systems have the same low albedo in visible ($p_v$ = 0.047 ± 0.003 based on IRAS catalog (Tedesco et al. 2002)), but possibly a different surface composition. Indeed, visible-near infrared spectroscopic analysis (Emery et al. 2011) showed that Hektor is redder (D-type in Tholen and Barucci, 1989 classification) than Patroclus (P-type). However, mid-IR emissivity spectra of these Trojan asteroids (Emery et al. 2006; Mueller et al. 2010) display similar features similar to C-type asteroids (Marchis et al. 2012). Variations in interior structure and composition could explain these differences and similarities in spectral and physical properties. To infer constraints on the target's grain density one needs to account for porosity. Indeed objects of this size are expected to preserve up to 50-60% porosity following accretion. Thermal evolution drives further compaction, which we model following the approach developed in Castillo-Rogez and Schmidt, 2010.

Temperatures achieved in (617) Patroclus and (624) Hektor remain too small to incur significant compaction – they remain below 120 K in both cases. This implies that Patroclus' interior should be a porous mixture of ice and rock with a grain density of about 1.6 ± 0.4 g/cm$^3$, assuming an average porosity of 50%. As (624) Hektor is bigger than (617) Patroclus, its deep interior could undergo some compaction so that the final internal structure is stratified in a low-porosity core (< 15%) and a porous icy shell that is 70 to 80 km thick (Figure 2D-E). This leads to a grain density of 2.0 ± 0.3 g/cm$^3$ for each component of the bilobed primary. We can rule out higher densities as they imply extreme assumptions on the amount of residual porosity in these objects. Hence, the two asteroids are characterized by grain densities in the same ballpark, yet there is 25% difference between the central density values inferred for each object.

## 4. CONCLUSION

By studying the long-term stability of the system orbit, we found that the system has not evolved significantly through tides. This suggests that the large eccentricity and inclination are primordial, and a remnant signature of the formation of this system. Interestingly, because the orbit is close to 1:10 and 2:21 orbit/spin resonances, a small change in the semi-major axis would have ejected the moon, or crashed it on the primary. It is therefore very likely that the moon's orbit has not varied since its formation. Also, because of the large inclination, the moon most probably did not accrete from a swarm of fragments after a catastrophic disruption, a scenario generally proposed for the formation of multiple asteroid systems (Michel et al. 2001; Durda, 2004). Instead we speculate that the moon could be an ejecta produced by the low-velocity encounter that formed the bilobed primary. After the encounter of the two components, the primary spun up faster, reaching a point of instability which led to the mass shedding (Descamps and Marchis, 2008). More extensive simulations are required to better understand the formation and evolution of this binary Trojan asteroid.

(624) Hektor's grain density is in the same ballpark as those measured at Kuiper belt objects, such as Triton (2.061 $g/cm^3$) and Pluto (2.03±0.06 $g/cm^3$). On the other hand a grain density of 1.6-1.7 $g/cm^3$ is typical of objects found in giant planet systems (e.g., uncompressed densities for Ganymede, Callisto, and Titan; average density of Uranus' satellites). Interestingly, Jupiter's moon Amalthea shows similar physical properties (size and density) with Patroclus (Anderson et al. 2005), yet, the large error bar on the grain density inferred for the latter body precludes any robust conclusion. It is important to note that (617) Patroclus and (624) Hektor exhibit different spectral properties in visible/near-infrared that were already interpreted as evidence for an origin of these objects in different reservoirs (Emery et al. 2011). (624) Hektor's red spectrum was associated with Centaurs while (617) Patroclus' less red spectrum was associated with a formation at 5 AU. (617) Patroclus would contain a large fraction of water ice, possibly clathrate hydrates, and few native organics. Our results further point to a different composition and origin for these trapped binary Trojan asteroids. However, refined shape and mass determinations are necessary in order to derive grain densities with accuracies sufficient to establish genetic links between these objects and other small body populations. Significant compositional similarities between the Trojans and giant planet moons would also place constraints on the amount of processing that took place within circumplanetary disks (Mosqueira and Estrada 2003). If the populations of asteroids orbiting in the Jupiter Trojan swarms are indeed sub-products of the migration of the giant planets (Morbidelli et al. 2005), their study will help understand the dynamically disruptive past of our solar system (Nesvorný et al. 2013). Our results reinforce the idea that the 5 to 15 AU region could have been an important source of primitive material to the inner Solar system.

Multiple Trojan asteroid systems, like (624) Hektor, and those not yet discovered, located at the crossroad between inner and outer small solar system body frontiers contain clues to understand the complex history of our solar system.

**Acknowledgements:** F.M. acknowledges the support of NASA grant NNX11AD62G. These data were obtained with the W. M. Keck Observatory, which is operated by the California

Institute of Technology, the University of California, Berkeley, and the National Aeronautics and Space Administration. The work of J.D. was supported by grant P209/10/0537 of the Czech Science Foundation. We would like to thank A. W. Harris for his constructive and useful comments, which significantly improved this manuscript. M. Cuk's work was supported by NASA OPR grant NNX11AM48G. Thanks to Hélène Marchis for drawing the figures 3D&E.

*Figure 1:* Nine adaptive optics K$_p$ and H bands observations of (624) Hektor binary system taken with W.M. Keck telescope in LGS (blue labels) and in NGS (green labels). The location of the moon is indicated with a green circle of 0.2" diameter. The moon is barely visible on some NGS data due to the poor AO correction at the epoch of the observations. The center of the image shows the high level of intensity of the image, hence the primary. Hektor's primary is elongated, resolved in one direction, and displays a bilobed shape in some of the best images.

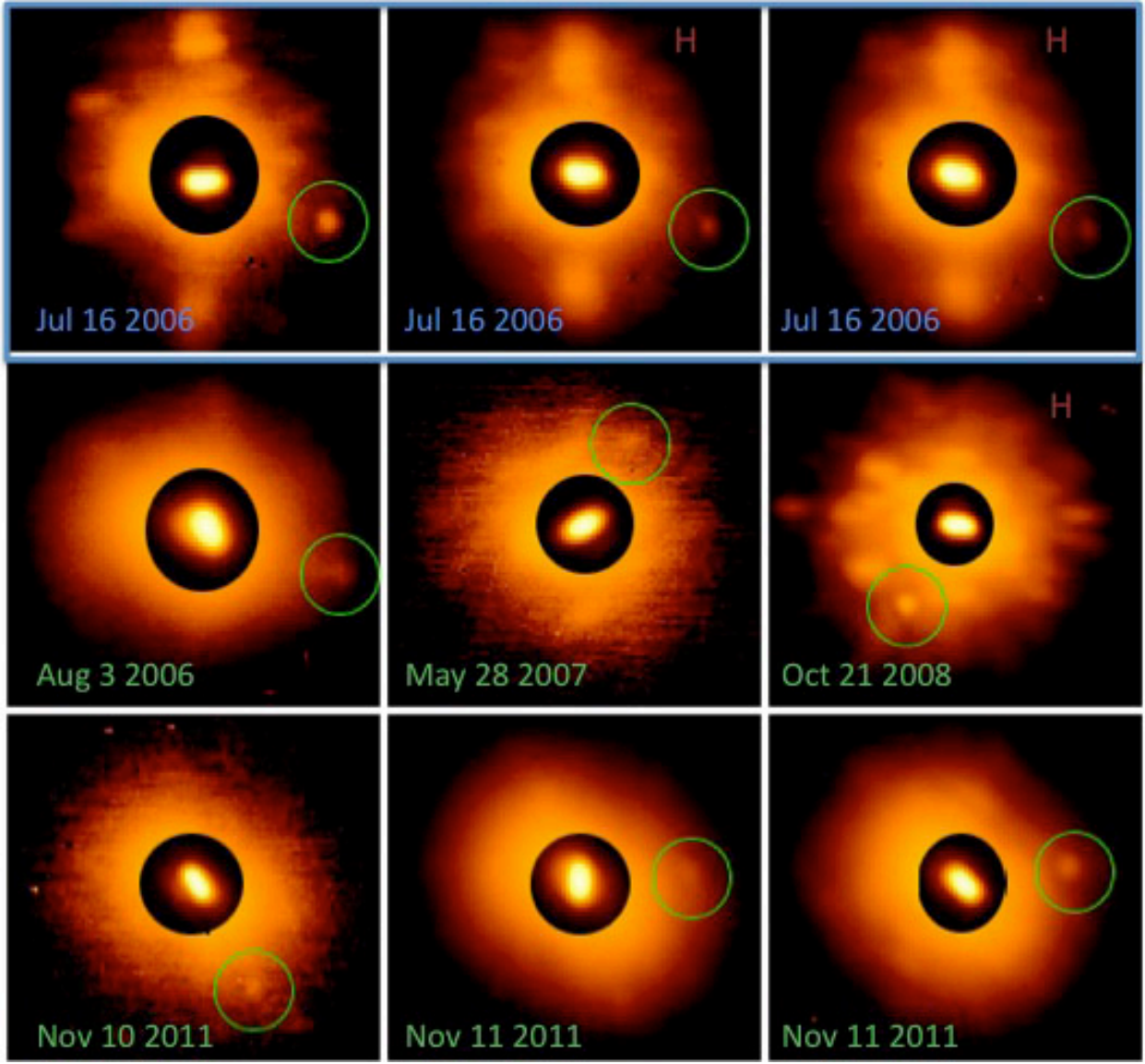

*Figure 2:* [A-C] Three 3D-shape models (polar view on the left, equator views in the middle and on the right) of Hektor's primary derived from 75 lightcurves collected since 1957. The convex, bilobed and one binary (or doublet) shapes have equal fitting quality. We rejected the binary shape since it poorly fits HST/FGS data. [D-E] Shape model of L4 Hektor's primary based on this work and L5 doublet (617) Patroclus-Menoetius based on previous works. The interiors of both binary Trojan systems are based on thermal evolution and compaction modeling. Hektor presents a stronger gradient in density due to the effect of internal pressure, which led to a core of compacted icy-rock material (porosity < 15%), whereas the outer layer remains porous (porosity ~50-60%). Owing to smaller internal pressures, the material forming Patroclus' component remained mostly uncompacted, making this binary asteroid a truly primitive Trojan.

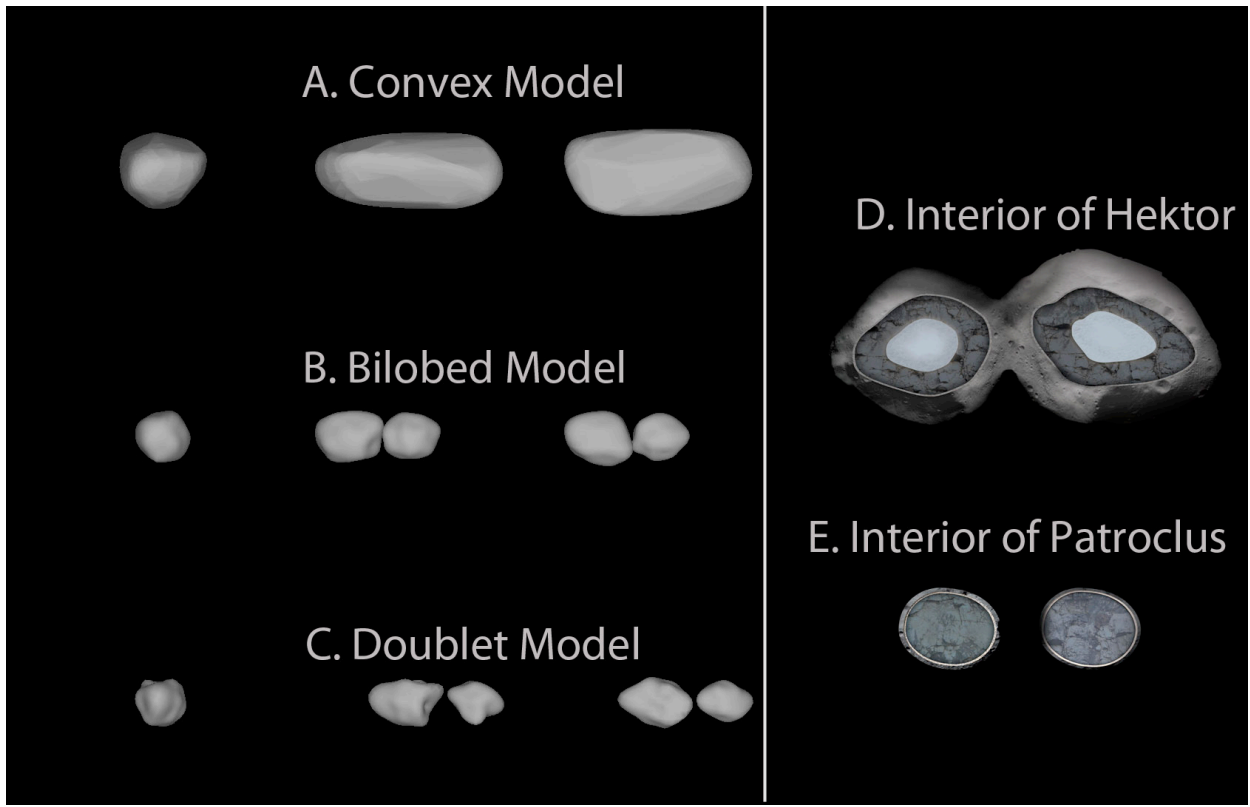

*Figure 3:* [A,B,C,D,F,J] Simulation of the nominal mutual orbit over one heliocentric orbit of Hektor. A symplectic integrator incorporating Hektor's $J_2$ and solar perturbations (Ćuk and Gladman, 2009) was used for all integrations shown here. [E] Inclination relative to Hektor's equator over 100 years, with the oscillation arising from proximity between the nodal precession period and half of Hektor's heliocentric period. [H,I] Evolution of the mutual orbit over 100,000 years.

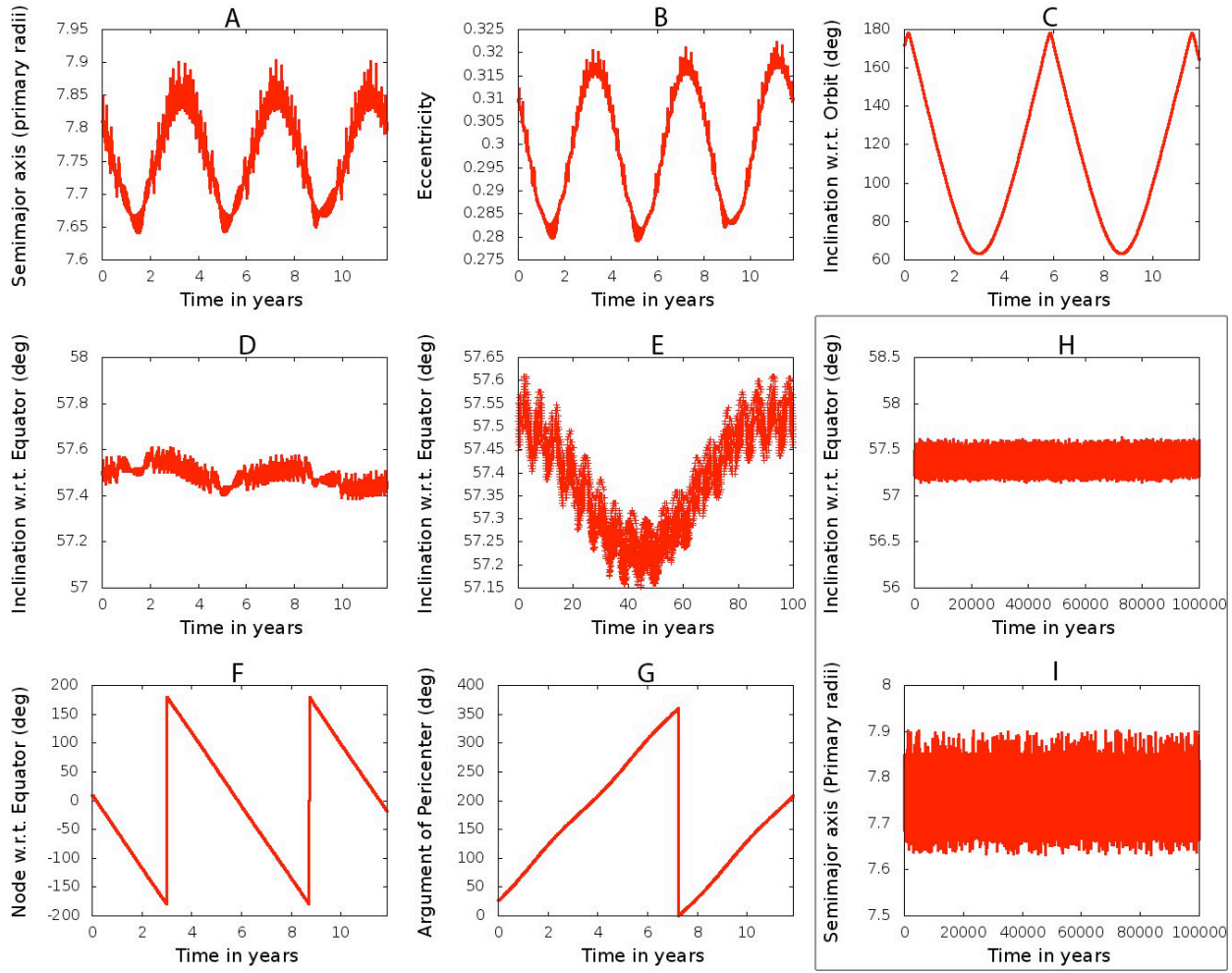

**Table 1**: *List of astrometric positions. $X/Y_{obs}$ are the astrometric positions relative to the primary in arcsec measured on the image in the East-West/North-South direction with a Moffat-Gauss fit. $X/Y_{calc}$ are the predicted positions from Genoide-ANIS dynamical model. $X/Y_{omc}$ are the difference. The average $X_{omc}$ is 1.8 milli-arcsec. The average $Y_{omc}$ is -5.4 milli-arcsec. The fitness value of the best solution is $f_p = 9.96$ mas*

| Julian Date | ISO date | Xobs arcsec | Yobs arcsec | Xcalc arcsec | Ycalc arcsec | Xomc arcsec | Yomc arcsec |
|---|---|---|---|---|---|---|---|
| 2453933.079 | 2006-07-16T13:53:29.996 | -0.3412 | -0.1216 | -0.3559 | -0.118 | 0.0148 | -0.0036 |
| 2453933.084 | 2006-07-16T14:00:34.997 | -0.3347 | -0.1125 | -0.3546 | -0.1185 | 0.0199 | 0.0061 |
| 2453933.087 | 2006-07-16T14:05:45.000 | -0.3452 | -0.1065 | -0.3535 | -0.1189 | 0.0083 | 0.0124 |
| 2453950.970 | 2006-08-03T11:16:22.995 | -0.3354 | -0.0817 | -0.3369 | -0.1067 | 0.0014 | 0.025 |
| 2453951.016 | 2006-08-03T12:22:58.002 | -0.3267 | -0.1093 | -0.3196 | -0.1117 | -0.0071 | 0.0024 |
| 2453951.048 | 2006-08-03T13:08:23.999 | -0.3147 | -0.1098 | -0.3068 | -0.1147 | -0.0079 | 0.005 |
| 2454249.084 | 2007-05-28T14:01:20.997 | -0.1362 | 0.1973 | -0.1247 | 0.2271 | -0.0115 | -0.0298 |
| 2454249.090 | 2007-05-28T14:08:59.003 | -0.1470 | 0.2037 | -0.1257 | 0.2283 | -0.0213 | -0.0246 |
| 2454760.953 | 2008-10-21T10:52:33.024 | 0.1304 | -0.1756 | 0.1217 | -0.1810 | 0.0087 | 0.0053 |
| 2454760.957 | 2008-10-21T10:57:46.656 | 0.1217 | -0.1822 | 0.1246 | -0.1795 | -0.0028 | -0.0027 |
| 2455876.130 | 2011-11-10T15:07:36.001 | -0.1491 | -0.2584 | -0.1438 | -0.2257 | -0.0053 | -0.0327 |
| 2455877.149 | 2011-11-11T15:34:20.303 | -0.2808 | 0.0422 | -0.2767 | 0.0580 | -0.0041 | -0.0158 |
| 2455879.156 | 2011-11-13T15:44:26.537 | -0.1864 | -0.2361 | -0.1700 | -0.2187 | -0.0164 | -0.0173 |

*Table 1:* Characteristics of (624) Hektor multiple system. The $J_2$ dynamical, and mass and orbital parameters were derived using Genoide-ANIS. We display the characteristics of the bilobed model, which this study found to provide the best match to the dataset.

| Component | Parameter | Value | 1-σ Error |
|---|---|---|---|
| **Primary** | Mass ($10^{18}$ kg) | 7.9 | 1.4 |
| | Equivalent Diameter Primary (km) | 250 | 26 |
| | Bulk density (g/cm$^3$) | 1.0 | 0.3 |
| | Dynamical $J_2$ | 0.15 | 0.04 |
| **Bilobed model** | Equivalent Diameter A (km) | 220 | 22 |
| | Equivalent Diameter B (km) | 183 | 18 |
| | Rotation period (hours) | 6.92050885 | 0.000002 |
| | Pole solution in ecliptic ECJ2000 (degrees) | λ=332 | 10 |
| | | β=-32 | 10 |
| | Theoretical $J_2$ | 0.10 | |
| **Moon** | Mass | Negligible | Set to 0 |
| | Estimated diameter (km) | 12 | 3 |
| | Semimajor axis (km) | 623.5 | 10 |
| | Period (days) | 2.9651 | 0.0003 |
| | Eccentricity | 0.31 | 0.03 |
| | Inclination in EQJ2000 (degrees) | 166.2 | 3.2 |
| | Inclination w.r.t. to the primary (degrees) | 50.1 | 1.1 |
| | Longitude of the Ascending node in EQJ2000 (degrees) | 170.7 | 6.1 |
| | Argument of periapsis in EQJ2000 (degrees) | 113.4 | 1.4 |
| | Pericenter date (days) | 2453928.287 | 0.071 |
| | Pole orientation in ecliptic ECJ2000 (degrees) | λ=282.8 | 4.1 |
| | | β=-80.0 | 3.3 |